\documentstyle[]{article}
\begin{document}
\begin{flushright}
{\bf IOP-TH-16/97\\
August, 1997}
\end{flushright}
\begin{center}
\vspace{2cm}

{\Large \bf The $b-\tau$ unification in GUTs with non-chiral matter}

\vspace{1cm}

A.B.~Kobakhidze

\vspace{0.5cm}

{\it Institute of Physics, GE-380077 Tbilisi, Tamarashvili Str.6,
Georgia}
\end{center}
\begin{abstract}
It is shown, that the presently accepted value for
$b$-quark mass can be obtained from the requirement of the exact
$b-\tau$ unification in the both non-SUSY and SUSY non-chiral
extended GUTs.
\end{abstract}

\vspace{1cm}

\paragraph{Introduction.}

In due time the minimal SU(5) grand unification proposal [1] had
ignited interest to the Grand Unified Theory (GUT) business by the
calculation of the unification scale $M_{GUT}$ and mixing angle
$sin^{2}\theta _{W}$ [2] and also by successful prediction of the
$b$-quark and $\tau $-lepton mass ratio, $R=m_b/m_{\tau}\approx 3$
[3,4].  It had been also shown, that the  predicted mass ratio
depends on the number of chiral quark-lepton families and that the
observed ratio seems to require three such families [4,5].  At
present we know that further prediction of three light families was
confirmed by determination of the number of neutrino species in $Z$-
boson decays at LEP and SLC [6]. However, we also know from the
present accurate data on the Standard Model (SM) gauge couplings
$\alpha _{S},\alpha _{W}$ and $\alpha _{Y}$, that the  minimal SU(5)
GUT, in which the first such predictions were made, is ruled out due
to the actual non-unification of the SM running gauge couplings. This
initiates one to go beyond the minimal SU(5) model.

One such way is a supersymmetric (SUSY) extension of the minimal
SU(5) GUT, in which gauge couplings meet each other at a single point
around $10^{16}GeV$ [7,8]. In addition to the unification of gauge
couplings, the unification of the $b$-quark and $\tau $-lepton Yukawa
couplings appears naturally [9].  Namely, in the small $tan\beta $
($tan\beta =v_{2}/v_{1}$ is the ratio of the two Higgs vacuum
expectation values (VEVs)) regime one obtains, that for the presently
allowed values of the electroweak parameters and the $b$-quark mass
$b-\tau $ unification demands for the large values of the $t$-quark
Yukawa coupling $Y_{t}=y_{t}^{2}/4\pi \simeq 0.1-1$ at the $M_{GUT}$
scale. These large values are exactly those that ensure the
attraction towards the infrared fixed point solution [10] of the
$t$-quark mass, providing an explanation for the heaviness of the
$t$-quark, $M_{t}=180\pm 12GeV$ [11]. From the other hand, although
there is exact $b-\tau $ unification in the large $tan\beta $ regime,
but as it was shown in [12], in this case the predicted $b$-quark
mass strongly depends on the SUSY particle spectra due to the
importance of the corrections induced by the SUSY breaking sector.

However, it is well known by now that the simple single scale
canonical SUSY GUTs predict the value of the strong gauge coupling (
$\alpha _{S}(M_{Z})\simeq 0.125$ or so) higher than the values
extracted from low-energy experiments ($\alpha _{S}(M_{Z})\simeq
0.11$) and the inclusion of threshold corrections do not change this
situation [13]. This discrepancy initiates many authors to consider
various extensions of the minimal SUSY GUT [14] or an alternative
string type unification [15].

Another way to achieve gauge coupling unification is the
split-multiplet non-SUSY SU(5) models first proposed by Frampton and
Glashow and extensively studied in refs [7,17]. It was shown, that
despite the gauge coupling unification most of such models with one
electroweak Higgs doublet predict too large value for the $b$-quark
mass, while in the case of two Higgs doublets the correct $b$-quark
mass could be achieved by setting $t$ -quark Yukawa coupling near its
infrared fixed point.

Needless to say, that the split-multiplet non-SUSY models are less
motivated than those of SUSY\ extended, mostly due to the gauge
hierarchy problem. However, until the experimental confirmation of
the SUSY they are real alternatives at least in the observable
aspects of unification.

In refs.[18,19] we have proposed a mechanism for the natural (without
any "fine tunings") explanation of the splitting of non-chiral
(vectorlike) multiplets in the frames of extended SU(2N) GUTs.  In
the minimal non-SUSY SU(6) version of such GUTs the natural
unification of the SM gauge couplings does appear assuming three
families of split fermions belonging to $15+\overline{15}$ SU(6) reps
with radiative induced masses [18]. In the case of SUSY SU(6) the
intermediate $G_{I}\equiv $ SU(3)$_{S}\otimes $ SU(3)$_{W}\otimes $
U(I) symmetry scale does exists naturally in the "Grand desert"
region and one family of such split superfilds is required in order
to achieve gauge coupling unification [19]. In the sharp contrast
with the minimal SUSY SU(5) case, this unification appears for the
low values of the strong gauge coupling as well and is close to the
string (or even Planck) unification which is welcomed feature from
the point of view of the string theories [20].

These attractive features of the non-chiral extended GUTs initiate us
to their further investigation. In the present paper we examine
non-SUSY and SUSY non-chiral extended SU(6) models of refs [18,19] by
the calculation of the $b-\tau $ mass ratio.

\paragraph{The $b-\tau $ unification in non-SUSY\ SU(6).}

Let us consider non-SUSY SU(6) GUT\ of ref. [18] with the fermion
content \begin{equation} 3\cdot (2\cdot \overline{6}+15)+N_{nc}\cdot
(15+\overline{15})~,  \label{1} \end{equation} where the first term
includes three chiral families of ordinary quarks and leptons while
the second one corresponds to the $N_{nc}$ non-chiral families of the
complementary fermions with the naturally light split submultiplets.
This later on the language of G$_{SM}\equiv $ SU(3)$_{S}\otimes $
SU(2)$_{W}\otimes $ U(1)$_{Y}$ decomposition are (see [18,19] for
more details ) :  \begin{equation} N_{nc}\cdot \left[
(3,2,\frac{1}{3})+(\overline 3,2,-\frac{1}{3})+ (3,1,-
\frac{2}{3})+(\overline 3,1,\frac{2}{3})\right]  \label{2}
\end{equation}

The mass $M_{SF}$ of split submultiplets (2) strongly depends on a
number of the starting non-chiral families of the complementary
fermions in order to provide the final unification. This easily can
be seen at Table 1 obtained by numerical integration of the two-loop
renormalization group equation (RGE) system for the running gauge
couplings\footnote{The $\beta $ functions below the energy $M_{SF}$
are those of two Higgs doublet SM model while those above the
$M_{SF}$ are modified by inclusion of the split fermion contributions
[18].}. As pointed in [18], the case of three ($N_{nc}=3$) families
of the complementary fermions looks quite natural if one assumes
radiative origin of the masses of split multiplets (2),
$M_{SF}\sim \alpha ^2_{GUT}\cdot M_{GUT}$.

The scalar sector besides the two heavy scalars $\Sigma \sim 35$ and
$ \varphi \sim 6$ (which break SU(6) down to the SM and provide
splitting between submultiplets of the complementary fermions)
includes also two Salam-Weinberg doublets from $H_1\sim 6$ and
$H_2\sim 15$ as it is usually in SU(6) in order to give masses to
$down$ and $up$ quarks, respectively, from the Yukawa interactions:
\begin{equation} y^{(down)} \overline {6}~ 15~ \overline {H}_1 +
y^{(up)} 15~ 15~ H_2 + h.c.  \label{3} \end{equation} From (3),
ignoring any mixing between families, we can write down the masses of
the $top$, $ bottom$ and $tau$, which are given by VEVs of $H_1$ and
$H_2$ as :  \begin{equation}
m_t=y_tvsin\beta,~~m_b=y_bvcos\beta,~~m_{\tau}=y_{\tau}vcos\beta~,
\label{4} \end{equation} where
$v=\frac{1}{\sqrt{2}}(H_1^2+H_2^2)^{1/2}$ and
$tan\beta=\frac{<H_2>}{<H_1>}$.  The appropriate one-loop RGEs for
$Y_t, Y_b$ and $Y_{\tau}$ ($Y_a=y^2_a/4\pi, a=t, b, \tau$) have the
form :  \begin{equation} \mu\frac{d\ln
Y_a}{d\mu}=\frac{1}{2\pi}\biggl (C_{ab}Y_b-D_{ai}\alpha _i\biggr )~,
\label{5} \end{equation} where $i$ runs over the $Y, W, S$ indices;
$\mu$ denotes the renormalization scale and the numerical factors
$C_{ab}, D_{ai}$ are those as in two Higgs doublet extended SM [21].

Taking the experimental data for gauge couplings and the physical
mass $t$-quark [11,22] \begin{eqnarray} \alpha _S(M_Z)=0.118\pm 0.005
\nonumber \\ sin^2\theta _W(M_Z)=0.2312\pm 0.0003  \nonumber \\
\alpha ^{-1} _{EM}(M_Z)=127.9\pm 0.2  \nonumber \\ M_t=180\pm 12 GeV
\label{6} \end{eqnarray} and using $b-\tau$ unification condition,
$R(M_{GUT})=1$, we calculate the $ b $-quark pole mass\footnote{
Two-loop QCD dressed pole mass relates with running mass $m(\mu)$ (4)
by the well known formulae : \\ $M=m(M)\biggl (1+\frac{4}{3\pi}\alpha
_S(M)+12.4\cdot (\frac{\alpha _S(M)}{\pi})^{1/2}\biggr )$} by the
numerical integration of the RGEs system (5).  We have presented the
results of our calculations as a dependence of the  $b$-quark pole
mass $M_b$ on the $tan\beta$ for small $tan\beta$ regime (Fig.1) as
well as for large $tan\beta$ regime (Fig.2), using the values of
strong gauge coupling $\alpha _S=0.11, 0.117$ and $M_t=180 GeV,
M_{\tau}=1.778 GeV$ for the $t$-quark and $\tau$-lepton masses,
respectively. Dashed lines in Figs.1,2 correspond to the case of one
family ($N_{nc}=1$) of the non-chiral fermions while solid lines to
the three family case ($N_{nc}=3$).

One can see from these Figs. that the predicted $b$-quark mass
decreases with increasing of the number of non-chiral families and
with decreasing of $\alpha _{S}(M_{Z})$. This tendency is quite
favorable because, as it was mentioned above, natural values for the
split fermion masses ($M_{SF}> \alpha _{GUT}^{2}M_{GUT}\sim
10^{12}GeV$) can be obtained only for $N_{nc}\geq 3$. So, we can
conclude that the predicted $b$-quark mass in non-chiral extended
non-SUSY SU(6) is in a good agreement with the observed value $
M_{b}=4.9-5.2GeV$ [22] for the small $tan\beta \simeq O(1-3)$ as well
as for the large $tan\beta \simeq O(40-60)$ and for the values of the
strong coupling constant $\alpha _{S}(M_{Z})\leq 0.12$.

\paragraph{The $b-\tau $ unification in SUSY\ SU(6).}

Now let us consider the SUSY extension of the model considered in the
previous section. The essential point related with SUSY extension
seems to be that the general superpotential of the Higgs superfields
$\Sigma \sim 35$, and $\varphi \sim 6+\overline{\varphi }\sim
\overline{6}$ allows SU(6) breaking mainly along the foregoing
$G_{I}=SU(3)_{S}\otimes SU(3)_{W}\otimes U(I)$ channel providing the
splitting of the $15+\overline{15}$ matter superfields just as in
non-SUSY case. The $G_{I}$ intermediate symmetry scale $M_{I}$ given
by VEVs of the $\varphi (\overline{\varphi })$ can be expressed
through the basic parameters of the model -- the unification scale
$M_{GUT}$ and the masses of split states (2) (now chiral superfields)
$M_{SF}$ as :  \begin{equation} M_{I}=\left[ M_{GUT}M_{SF}\right]
^{1/2}\eta , \end{equation} where $\eta \sim O(1)$ is the
dimensionless parameter expressed through the coupling constants of
$\Sigma $, $\varphi (\overline{\varphi })$ and $15+\overline{15}$
superfield interactions [19]. While in non-SUSY\ case the natural gap
between masses $M_{SF}$ and $M_{GUT}$ would be at most the radiative
one, in SUSY case the mass scale $M_{SF}$ in principal could be much
lower and even down to the SUSY breaking scale. In Table 2 we present
the predictions of the $M_{SF}$ for the one family of $15+\overline
{15}$. This predictions are based on the requirement of the gauge
coupling unification at two-loop level, using as an effective SUSY
scale $T_{SUSY}=M_{Z}$ [23], M$_{t}=180GeV$, $\sin \theta
_{W}=0.2312$, $\alpha _{EM}^{-1}=127.9$ and $\alpha _{S}=0.11$ and
0.125.

One can see from Table 2, that in a sharp contrast with the canonical
SUSY GUT situation the influence of the new gauge interactions below
the intermediate scale $M_I$ together with the contribution coming
from the split states (2) provide the increasing of the unification
scale up to the string M$_{GUT}\simeq $M$_{string}\simeq 5.5\cdot
10^{17}g_{GUT}$ GeV and even the Planck scale. This is a welcomed
feature from the point of view of string theories [20] and on the
other hand such large unification scale gives the sufficient
suppression of the $d=5$ operator induced proton decay for the whole
range of the $\tan \beta $ parameter.

Now let us look what happens with $b-\tau $ unification. Besides the
Higgs superfields $\Sigma $ and $\varphi (\overline \varphi )$ as in
non-SUSY case we introduced the additional Higgs superfields
$H_{1}\sim 6+\overline{H}_{1}\sim \overline{6}$ and $H_{2}\sim
15+\overline{H}_{2}\sim \overline{15}$ and by appropriate fine tuning
of the tree level superpotential parameters extract Salam-Weinberg
doublet and antidoublet from $H_{2}$ and $\overline{H}_{1}$
respectively. Assuming effective gauge theory for any given energy
region one can calculate $\beta $-functions for gauge coupling RGEs
as well as numerical factors $C_{ab}$ and $D_{ai}$ for the Yukawa
couplings (5) [21]. For the energy region below the intermediate
scale $M_I$ the numerical factors $C_{ab}$ and $D_{ai}$ are exactly
those as in minimal SUSY model [21] while are significantly modified
above the $M_I$ :  \begin{equation} C_{ab}=\left( \begin{array}{lll}
7 & 1 & 0 \\ 1 & 7 & 1 \\ 0 & 3 & 3 \end{array} \right)
,~~~D_{ai}=\left( \begin{array}{lll} \frac{16}{3} & \frac{16}{3} &
\frac{4}{3} \\ \frac{16}{3} & \frac{16}{3} & \frac{2}{3} \\ 0 & 8 & 1
\end{array} \right)   \label{7} \end{equation}

The results of numerical integration of RGEs (5) with appropriate
numerical factors as in the previous section are presented
graphically in the $M_{b}-\tan \beta $ plane for the gauge coupling
unification solutions from Table 2 and for small and large
$\tan\beta $ regimes respectively at Fig.3 and Fig.4. Note that the
influence of new gauge and Yukawa interactions leads to the
decreasing of $M_{b}$ in the both small and large $\tan \beta $
cases. As it is well known, the decreasing of the strong gauge
coupling works in the same direction, so to obtain the correct value
for $b$-quark mass one can take the values of $t$-quark Yukawa
coupling at $M_{GUT}$ significantly lower than in the case of
canonical SUSY SU(5). This shifts the prediction of the $t$-quark
mass from its infrared fixed point.

To conclude, I have shown that the presently accepted value for
$b$-quark mass can be obtained from the requirement of the exact
$b-\tau $ unification in the both non-SUSY and SUSY non-chiral
extended GUTs. Namely, in non-SUSY case the small and large $\tan
\beta $ regimes demand low values of the strong gauge coupling
$\alpha _{S}<0.12$ for the tree non-chiral families and $\alpha
_{S}<0.116$ for one family of such fermions. In the SUSY case, the
influence of new gauge and Yukawa interactions decrease the $b$-quark
mass value relative to the standard SU(5) situation and shift the
$t$-quark mass from its infrared fixed point.

\vspace{1cm}

\section*{Acknowledgement}

Many useful discussions with J.Chkareuli and I.Gogoladze is greatly
acknowledged. I would like also to thank A.Barnaveli for reading the
manuscript. This work was partially supported by the Grant No.2.10 of
the Georgian Academy of Sciences and the INTAS Grant RFBR 95-567.

\newpage

\begin{table}[h] \caption{The dependence of the mass of split
submultiplets $M_{SF}$ on the number of non-chiral families $N_{nc}$
and $\alpha _S(M_Z)$ obtained from the requirement of gauge coupling
unification at two-loop level in non-SUSY SU(6).}

\vspace{1cm}

\begin{center} 
\begin{tabular}{ccccc}\hline \hline
~$N_{nc}$~&~$\alpha _S(M_Z)$~&~$M_{SF}[GeV]$~&~$M_{GUT}[GeV]$~&~
$\alpha ^{-1}_{GUT}$~  \\ & & & & \\ & 0.11 &$10^{5.19}$ &
$10^{15.55}$ & $36.35$ \\ 1    & 0.117&$10^{3.87}$ & $10^{15.95}$ &
$35.46$ \\ & 0.125& $10^{2.60}$ &$10^{16.35}$ & $34.56$ \\ & & & & \\
     & 0.11 &$10^{12.05}$ & $10^{15.55}$ & $36.35$ \\ 3    &
0.117&$10^{11.91}$ & $10^{15.90}$ & $35.59$ \\ & 0.125&
     $10^{11.76}$&$10^{16.25}$ &  $34.80$\\ \hline \hline
\end{tabular} \end{center} \end{table}

\vspace{3cm}

\begin{table}[h] \caption{The representative solutions of the
two-loop RGEs for gauge couplings in non-chiral extended SUSY SU(6).}

\vspace{1cm}

\begin{center} %
\end{center}
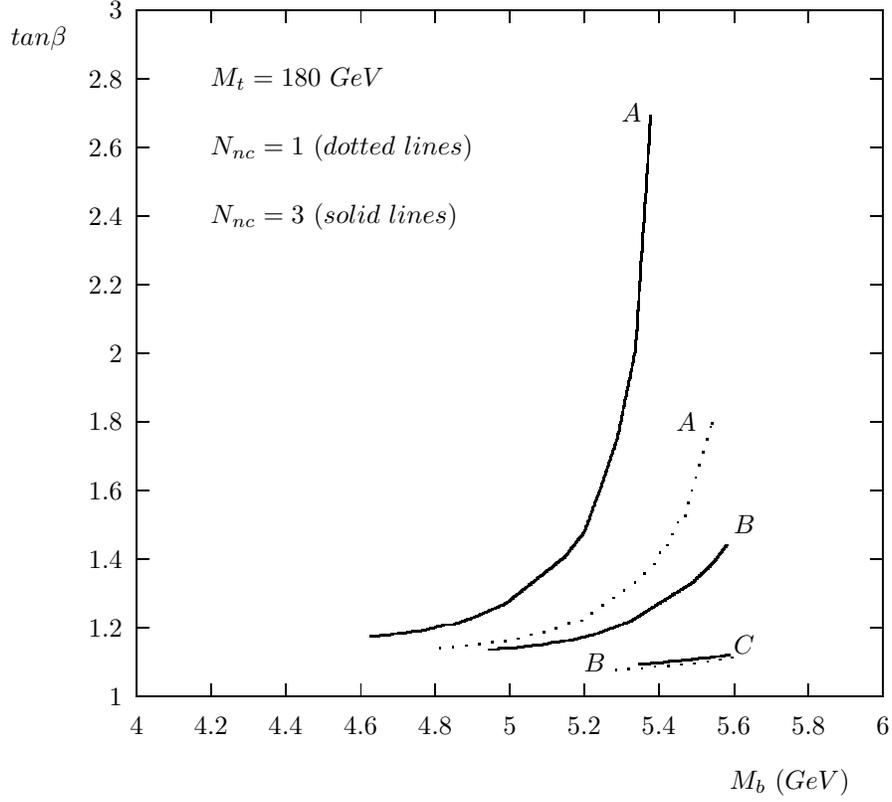
\caption{The $b$-quark mass as a function of $tan
\beta$ (small $tan\beta$ regime) in non-SUSY SU(6) model with
non-chiral split fermions. The solid lines correspond to the case of
three family ($N_{nc}=3$) of non-chiral fermions, while those of
dotted to the one family case ($N_{nc}=1$). Lines denoted by A, B, C
correspond to the values of $\alpha _S(M_Z)=0.11,~0.117$ and 0.122,
respectively.}
\end{figure}

\newpage

\vspace{3cm}

\begin{figure}[h]
\begin{center}
\setlength{\unitlength}{0.240900pt}
\ifx\plotpoint\undefined\newsavebox{\plotpoint}\fi
\sbox{\plotpoint}{\rule[-0.175pt]{0.350pt}{0.350pt}}%

\end{center}
\caption{The same as in Fig.1 in the large $\tan\beta$
regime.}
\end{figure}

\newpage

\begin{figure}[h]
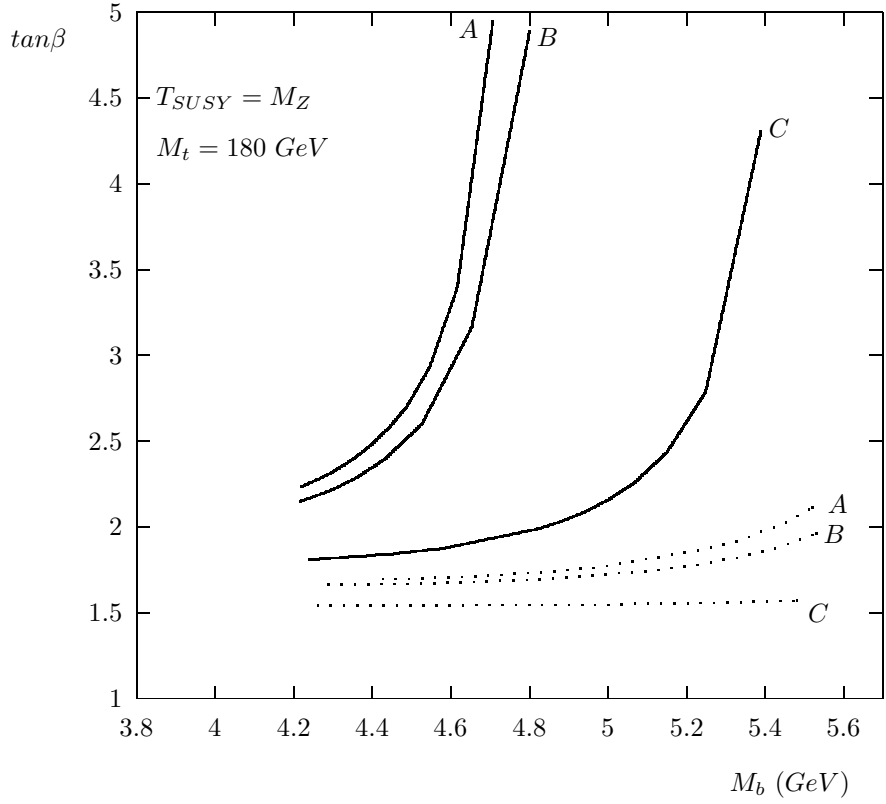

\begin{center}
\setlength{\unitlength}{0.240900pt}
\ifx\plotpoint\undefined\newsavebox{\plotpoint}\fi
\sbox{\plotpoint}{\rule[-0.175pt]{0.350pt}{0.350pt}}%
%
\end{center}
\caption{The $b$-quark mass as a function of $tan
\beta$ (small $tan\beta$ regime) in SUSY SU(6) model. The solid lines
correspond to the value of $\alpha _S(M_Z)=0.11$, while those of
dotted to the value of $\alpha _S(M_Z)=0.122$. Lines denoted by A, B,
C correspond to A, B, C cases of gauge coupling unification (see
Table 2).}
\end{figure}

\newpage

\vspace{3cm}

\begin{figure}[h]
\begin{center}
\setlength{\unitlength}{0.240900pt}
\ifx\plotpoint\undefined\newsavebox{\plotpoint}\fi
\sbox{\plotpoint}{\rule[-0.175pt]{0.350pt}{0.350pt}}%
%
\end{center}
\caption{The same as in Fig.3 in the large $tan\beta$
regime.}
\end{figure}


\begin{thebibliography}{99} \bibitem{} H.Georgi and S.L.Glashow,
Phys.Rev.Lett. 32 (1974) 438.  \bibitem{} H.Georgi, H.R.Quinn and
S.Weinberg, Phys.Rev.Lett. 33 (1974) 451.  \bibitem{} H.S.Chanowitz,
J.Ellis and M.K.Gaillard, Nucl.Phys. B126 (1977) 506.  \bibitem{}
A.J.Buras, J.Ellis, M.K.Gailard  and D.V.Nanopoulos, Nucl.Phys. B135
(1978) 66.  \bibitem{} D.V.Nanopoulos and D.A.Ross, Nucl.Phys. B157
(1979) 273; Phys.Lett.  B108 (1982) 351.  \bibitem{} ALEPH Coll.,
D.Decamp et al., Phys.Lett.  B239 (1989) 519; DELPHI Coll., P.Aarnio
et al., $ibid$ 539;  L3 Coll., B.Adeva et al., $ibid$ 509;  OPAL
Coll., M.Z.Akrawy et al., $ibid$ 530; \\ Mark II Coll., G.S.Abrams et
al., Phys.Rev.Lett. 63 (1989) 724; $ibid$ 2173.  \bibitem{} U.Amaldi
et al., Phys.Lett. B260 (1991) 131; B281 (1992) 374.  \bibitem{}
C.Giunti, C.W.Kim and U.W.Lee, Mod.Phys.Lett. A6 (1991) 1745 \\
J.Ellis, S.Kelly and D.V.Nanopoulos, Phys.Lett. B260 (1991) 441 \\
P.Langacker and M.Luo, Phys.Rev. D44 (1991) 817.  \bibitem{} H.Arason
et al., Phys.Rev.Lett. 67 (1991) 2933; \\ S.Kelly,  J.Lopez and
D.V.Nanopoulos, Phys.Lett. B278 (1992) 140; \\ S.Dimopoulos, L.Hall
and S.Raby, Phys.Rev. D45 (1992) 4192; \\ V.Barger, M.S.Berger and
P.Ohman, Phys.Rev. D47 (1993) 1093; \\ V.Barger, M.S.Berger, P.Ohman
and R.J.N.Phillips, Phys.Lett. B314 (1993) 351; \\ P.Langacker and
N.Polonsky, Phys.Rev. D49 (1994) 1454; \\ W.A.Bardeen, M.Carena,
S.Pokorski and C.E.M.Wagner, Phys.Lett. B320 (1994) 110.  \bibitem{}
B.Pendleton and G.G.Ross, Phys.Lett. B98 (1981) 291; \\ C.T.Hill,
Phys.Rev. D24 (1981) 691.  \bibitem{} CDF Coll., F.Abe et al.,
Phys.Rev. D50  (1994) 2966; Phys.Rev.Lett.  73 (1994) 225; $ibid$ 74
(1995) 2226; \\ D0 Coll.,S.Abachi et al.,  Phys.Rev.Lett. 72 (1994)
2138; $ibid$ 74 (1995) 2632.  \bibitem{} R.Hempfling, Phys.Rev. D49
(1994) 6168; \\ L.J.Hall, R.Ratazzi and U.Sarid, Phys.Rev. D50 (1994)
7048; \\ M.Carena, M.Olechowski, S.Pokorski and C.E.M.Wagner,
Nucl.Phys. B426 (1994) 269.  \bibitem{} J.Barger, K.Matchev and
D.Pierce, Phys.Lett. B348 (1995) 443; \\ R.H.Chankowski,
Z.Pluciennik, S.Pokorski and C.E.Vyonakis, Phys.Lett.  B358 (1995)
264; \\ R.Barbieri, P.Clafaloni and A.Strumia, Nucl.Phys.  B442
(1995) 461.  \bibitem{} B.Brachmachari and R.N.Mohapatra, Phys.Lett.
B357 (1995) 566; \\ J.Ellis, J.L.Lopez and D.V.Nanopoulos, preprint
CERN-TH-95/260, hep-ph/9510246; \bibitem{} K.Dienes and A.Farragi,
Phys.Rev.Lett.  75 (1995) 2646; Nucl.Phys.  B457 (1995) 409; \\
L.Roszkowski and M.Shifman, Phys.Rev D53 (1996) 404.  \bibitem{}
P.H.Frampton and S.L.Glashow, Phys.Lett. B131 (1983) 340(E); B135
(1984) 515.\\ \bibitem{} S.Nandi, Phys.Lett. B142 (1984); \\
H.Murayama and T.Yanagida, Tohoku University preprint TU-370 (May,
1991); \\ A.Giveon, L.J.Hall and U.Sarid, LBL preprint LBL-31084
(July, 1991); \\ P.H.Frampton, J.T.Liu and M.Yamaguchi, Phys.Lett.
B277 (1992) 130.  \bibitem{} J.L.Chkareuli, I.G.Gogoladze and
A.B.Kobakhidze,  Phys.Lett. B340 (1994) 63.  \bibitem{}
J.L.Chkareuli, I.G.Gogoladze and  A.B.Kobakhidze, Phys.Lett.  B376
(1996) 111.  \bibitem{} V.Kaplunovsky, Nucl.Phys.  B307 (1988) 145;
\\ I.Antonadis, J.Ellis, R.Lacaze and D.V.Nanopoulos, Phys.Lett. B268
(1991) 188; \\ K.Dienes and A.Faraggi, see ref. 15.  \bibitem{}
M.E.Machacek and M.T.Vaughn, Nucl.Phys. B222 (1983) 83; $ibid$ B236
(1984) 221. \bibitem{} Particle Data Group, Phys.Rev.  D50 (1994)
1173.  \bibitem{} P.Langacker and N.Polonsky, Phys.Rev.  D47 (1993)
4028; \\ M.Carena, S.Pokorski and C.E.M.Wagner, Nucl.Phys.  B406
(1993) 59.  \end{thebibliography}
\end{document}